\documentclass[conference]{IEEEtran}
 
\ifCLASSINFOpdf
\else
\fi
\usepackage{amssymb}
\usepackage{amsmath}
\usepackage[linesnumbered,ruled]{algorithm2e}
\usepackage{graphicx}
\graphicspath{ {./images/} } 
\usepackage{caption}
\usepackage{subcaption}
\usepackage[english]{babel}
\usepackage[utf8]{inputenc}
\usepackage[nottoc]{tocbibind}
\usepackage{cite}
\begin{document}
%
\title{Dynamic Graph Learning based on Graph Laplacian}

\author{\IEEEauthorblockN{Bo Jiang$^1$, Ashkan Panahi$^2$, Hamid Krim$^1$,Yiyi Yu$^3$, Spencer L. Smith$^3$}
\IEEEauthorblockA{$^1$Department of Electrical and Computer Engineering, North Carolina State University, Raleigh, NC, USA\\
$^2$Department of Computer Science and Engineering, Chalmers University of Technology, Göteborg, Sweden\\
$^3$Department of Electrical and Computer Engineering, University of California Santa Barbara, Santa Barbara, CA, USA\\
bjiang8@ncsu.edu}}


%


\maketitle

\begin{abstract}
The purpose of this paper is to infer a global (collective) model of  time-varying responses of a set of nodes as a dynamic graph, where the individual time series are  respectively observed at  each of the nodes. The motivation of this work lies in the search for a connectome model which properly captures  brain functionality upon  observing activities in different regions of the brain and possibly of  individual neurons. We formulate the problem as a quadratic objective  functional of observed node signals over short time intervals, subjected to the proper regularization reflecting the graph smoothness and other dynamics involving the underlying graph's Laplacian, as well as the time evolution smoothness  of the underlying graph. The resulting joint optimization  is solved by a continuous relaxation and an introduced novel gradient-projection scheme. We apply our algorithm to a real-world dataset comprising recorded activities of individual brain cells. The resulting model is shown to not only be viable but also efficiently computable.\\
\end{abstract}

%
\IEEEpeerreviewmaketitle

\section{Introduction}
\indent The increased and ubiquitous emergence of graphs is becoming an excellent tool for quantifying interaction  between different elements in great variety of network systems. Analyzing and discovering the underlying structure of a graph for a given data set has become central to a variety of different signal processing research problems which may be interpreted as graph structure recovery. For example, in social media \cite{jagadish2014discovery}, such as Facebook and LinkedIn, the basic interaction/relation between two individuals being represented by a link, yields the notion of a graph known as the social network, which is used for inferring further characteristics among all involved individuals \cite{huang2018fusion}. Similarly, in physics \cite{audi2003ame2003} and chemistry \cite{canutescu2003graph}, graphs are widely used to capture the interactions among atoms/molecules to study the behavior of different materials. The rather recent connectome paradigm\cite{sporns2012discovering} in neuros-science, is based on the hypothesis that the brain is a massively connected network and its behavior variation and connectivity, particularly in response to controlled external, can be used to investigate brain's structures and ideally its functionality \cite{polania2011introducing, ocker2017statistics}.\\
\indent Existing  analysis approaches of connectivity of  neuron signals  and associated problems can be categorized into two main groups, $(i)$ Noise correlation analysis\cite{yu2018mesoscale, sompolinsky2001population}, which is often applied by neuro- scientists to uncover the connectivity among neurons, $(ii)$ Static graph learning \cite{maretic2017graph, chepuri2017learning}, which, by way of an optimization procedure, tries to attain a fixed graph over time. Noise correlation is commonly used  in neuro-science to establish connectivity between every pair of  neurons if their noise correlation  over a short time is significant.
  This method, however, requires many observations, making it hard to obtain an acceptable connectivity estimation over that interval. Additionally the acquired connectivity does not lend to simple rationalization following an experiment with a specific stimilus.
 In the second track, research on estimating graph structures for a given data set has been active  and includes graph learning. Research on the latter 
has primarily been  based on graphs' topology and signals' smoothness, and the application of the graph Laplacian has been predominant. 
Other recent work includes deep neural network modeling, and the training/testing was performed  on graph datasets to ultimately generate a graph to represent  patterns under given signals. These studies have primarily focussed on a static graph, with  signals non-sparse and the assumption of consistency of the graph over time. These models require sufficiently adequate samples for training and testing, once again making difficult to use on neuronal signals with a typically low sample size, in order to glean the desired variations over small time intervals.
Graphs' dynamics with clear potential impact on temporal data characterization, have also been of interest by many researchers \cite{goyal2019dyngraph2vec, goyal2018dyngem}. In this theme,  the models are used to predict the links given  previous graphs and associated signals. All these approaches require much prior information on known structures and plenty of data for training and predicting future graphs.\\
\indent Building on the wealth of prior research work in neuro-science and graph learning  \cite{chepuri2017learning}, we propose a new model, with a dynamic structure goal  to track neurons' dynamic connectivity over time. To that end, our proposed graph will include vertices/nodes for nodes,  and  their  connectivity is reflected by the graph  edges whose attribute is  determined by  the probability/intensity of connection between every pair of neurons.\\
To proceed, the outline of our paper follows in order our contributions in the sequel. Firstly, exploiting the insight from prior  research on graph learning with graph Laplacian \cite{chepuri2017learning, dong2016learning}, we propose an optimization formulation, which can yield  a graph over  each short time interval which in turn reflects the evolving transformation of  the connectivity. Secondly, we modify our model to fit sparse signals so that we can verify our optimized solution on a  neuronal signal dataset. Thirdly, we apply three alternative methods to simplify the optimization procedures, to simplify the solution procedures of the optimization problems, and help reach  the optimal points. We finally proceed to  test our proposed model on a neuronal dataset, to help improve our  understanding on the neuronal interaction and their process of transferring  signals.

\section{Problem Setup and Background}
\indent For notation clarity and throughout the paper, we will adopt  an upper and lower case bold letter to respectively denote a matrix  and a vector, and the  superscripts $T, -1$ to respectively denote its tranpose and inverse.  The operator $tr (\cdot)$ denotes a matrix trace.  The identity, zero and "1" matrices are respectively denoted by 
  $\mathbf{I}$, $\mathbf{0}$ and $\mathbf{1}$, while  $x_{ij}$ represents the $i$-th row, $j$-th column element of $\mathbf{X}$. \\
\indent Our neuronal-activity dataset will consist of $\mathnormal{N}$ neuron/nodes, and will be characterized by a connectivity graph $\mathnormal{G}=\{V,E,W\}$, where $\mathnormal{V}$ denotes the vertex set $V=\{v_1,v_2,\dots,v_N\}$,  $\mathnormal{E}$ is the edge set with attributes reflecting  the connectivity between each pair of vertices quantified by  $0\leq w\leq 1,\forall w\in W$ as a connectivity strength.  A time series $y_n(t)$ of observations with $t=1,2,\ldots,T$, is associated with each node $v_n$. For simplicity in our derivations, we will aggregate the nodes' finite length time series into a  $N\times T$ matrix $\mathbf{Y}$, where $(\mathbf{Y})_{n,t}=y_n(t)$. Our problem formulation will seek for each observed $\mathbf{Y}$, either a static  graph $G$ or a time dependent graph series of  graphs $G_1,G_2,\ldots$.\\ 
\indent The well known graph Laplacian of an undirected graph can  describe its topology, and can serve  as the second derivative operator of the underlying graph. The corresponding Laplacian matrix $\mathcal{L}$ is commonly defined as \cite{chung1997spectral},  with $\mathcal{L}(i,j)= -w_ij$, for  $i$,$j$ adjacent nodes, and $\mathcal{L}(i,j)= 0$ otherwise, and  $\mathcal{L}(i,i)= d_i$ where  $d_i=\sum_{j}w_{ij}$ denotes the degree of node $i$ .  Its simple matrix expression is  $\mathcal{L}=D-W$, where $\mathnormal{D}$ is a diagnal matrix of degrees.\\
\indent The Laplacian matrix may also usefully adopt, in some context,  a second derivative interpretation of graphs: Given an assignment $x=(x_1,x_2,\ldots,x_n)$ of real number to the nodes, the Laplacian matrix may be found as the second derivative of $x$ as  $\mathcal{L}(\mathbf{W},x)=\sum_i\sum_{j>i}w_{ij}\mathbf{a}_{ij}\mathbf{a}^T_{ij}x$,
where $\mathbf{a}_{ij}$ denotes a N-dimensional vector whose elements are all 0s, except the $i$-th element being respectively 1 and $j$-th element -1. As  may be seen, $a_{ij}$ represents the first derivative of the edge between the $i$-th and $j$-th node.
The notion of a Laplacian will be exploited in this sequel as  a structural regularizer when an optimal graph is sought for a given data set.

\section{Dynamic Graph Learning}

\subsection{Static Graph Learning }
\indent Prior to proposing the   dynamic structure learning of a graph, we first recall the principles upon which static graph learning was based \cite{chepuri2017learning}. Using the Laplacian quadratic form, $\mathbf{x}_n\mathcal{L}(\mathbf{W})\mathbf{x}_n$,  as a smoothness regularizer of the signals $\mathbf{x}_n$, and  the degree of connectivity $K$ as a tuning parameter, \cite{chepuri2017learning} discovers a $K$-sparse graph from noisy signals $\mathbf{y}_t=\mathbf{x}_t+\mathbf{n}_t$, by seeking the solution of the following, 
\begin{equation}
\begin{split}
    \operatorname*{argmin}_{\mathbf{X},\mathbf{W}} &\frac{1}{T}\sum_{t=0}^{T-1}\|\mathbf{y}_t-\mathbf{x}_t\|^2+\gamma\mathbf{x}_t^T\mathcal{L}(\mathbf{W})\mathbf{x}_t\\
    s.t. & \quad0\leq w_{t,ij}\leq 1,\quad\forall i,j, \quad\sum_{i,j>i}w_{t,ij}=K,
\end{split}
\end{equation}
where $\gamma$ and $K$ are tuning parameters, $\mathbf{X}=[\mathbf{x}_1,\mathbf{x}_2,\dots,\mathbf{x}_T]$ is the noiseless signals and $\mathbf{Y}=[\mathbf{y}_1,\mathbf{y}_2,\dots,\mathbf{y}_T]$ its noisy observation.  $\mathbf{W}$ is the adjacency weight matrix for the undirected graph, with the additional relaxation of the individual weights to the interval $[0,\,1]$. 
 
\subsection{Dynamic Graph Learning }
\indent Note that in \cite{chepuri2017learning}  a single connectivity graph is inferred for the entire observation time interval, thus overlooking the practically varying connections between every two nodes over time. To account for these variations and towards capturing the true underlying structure of the graph, we propose to learn the dynamics of the graph. Relating these dynamics to the brain signals which are of practical interest, they would not only reflect the signals (as a response to  the corresponding stimuli) in that time interval, but also account for their dependence on those in the previous graph and time interval. To that end, we can account for the similarity of temporally adjacent graphs in the overall functionality of the sequence of graphs consistent with the observed data. Selecting  a 1-norm distance of connectivity weight matrices between consecutive time intervals, we can proceed with the graph sequence discovery and hence the dynamics by seeking the solution to the following, 
\begin{equation}
\begin{split}
    \operatorname*{argmin}_{\mathbf{X},\mathbf{W}_t} &\sum_{t=1}^{T}\big[\|\mathbf{y}_t -\mathbf{x}_t\|^2  +tr(\gamma\mathbf{x}_t^T\mathcal{L}(\mathbf{W}_t)\mathbf{x}_t)\big]\\
    & +\alpha\sum_{t=1}^{T-1}\|\mathbf{W}_t-\mathbf{W}_{t+1}\|_1\\
    s.t. & \quad0\leq w_{t,ij}\leq 1,\quad\forall i,j, \quad\sum_{i,j>i}w_{t,ij}=K
\end{split}
\end{equation}
where $\alpha$ is the penalty coefficient, $\mathbf{Y}$ is the observed data, $\mathbf{X}=[\mathbf{x}_1,\mathbf{x}_2,\dots,\mathbf{x}_T]$ is the noise-free data, $\mathbf{W}_t$ is the weight matrix in the $t$-th time interval, and $K$ is a tuning parameter.

\subsection{Dynamic Graph Learning from Sparse Signals}
\indent The solution given by \cite{chepuri2017learning} addresses the static graph  learning, but the observed signals $y_n(t)$ may often be sparse, which poses a problem: Noting that  $\mathcal{L}(\mathbf{W},x)=\sum_i\sum_{j>i}w_{ij}\mathbf{a}_{ij}\mathbf{a}^T_{ij}x$ into the Laplacian quadratic form, we have $w_{ij}\|\mathbf{y}_i-\mathbf{y}_j\|^2$, calculating the distance between two signals, and we minimize this term to find some nodes with tough connections, in another word, the values of the signals are similar in a small time interval. This equation also implies that if $\mathbf{x}_i$ and $\mathbf{x}_j$ are close to 0, their distance will also be close to  0.  This thus introduces unexpected false edges when sparse signals are present. As a simple illustration, let us assume that sparse signals are rewritten as  $\mathbf{Y}=[\Tilde{\mathbf{Y}},\mathbf{0}]^T$, where the dimension of $\mathbf{Y}$ is $N\times t$, $\Tilde{\mathbf{Y}}$ is an $n\times t$ matrix and $\mathbf{0}$ is $(n-N)\times t$. Given that 2-norm is non-negative and the Laplacian matrix is positive semi-definite,  we can find a trivial optimal solution of $(\mathbf{X},\mathbf{W})$, where $\mathbf{W}$ is sparse, such that $\mathbf{X}=\mathbf{Y}$, and the weight matrix can be represented by some block matrix, $\mathbf{W}=\begin{bmatrix} \mathbf{0} & \mathbf{0}\\ \mathbf{0} & \mathbf{\Tilde{W}}\end{bmatrix}$.\\
\indent Since problem (3) is a convex and non-negative problem, it can be shown  that the optimal loss value is 0 by inserting the solution $\mathbf{Y}=\mathbf{X}$ and $\mathbf{W}$ into optimization (3).  This then shows that if sparse signals (which happens to be the case for brain firing patterns) are present, the solution to the formulated optimization may not be  unique , furthermore, yielding some of these optimal points to result in connections between zeros-signal nodes (i.e.  meaningless connections per  our understanding). \\
\indent Towards mitigating this difficulty, we introduce  a constraint term to help focus on the nodes with significant values, specifically we constrain edge nodes energy to be of significance. This yields the following formulation,
\begin{equation}
\begin{split}
    & \operatorname*{argmin}_{\mathbf{X},\mathbf{W}_t} \sum_{t=1}^{T}   \bigg[\|\mathbf{y}_t -\mathbf{x}_t\|^2  +tr(\gamma\mathbf{x}_t^T\mathcal{L}(\mathbf{W}_t)\mathbf{x}_t)\\
    &-2\eta\sum_{i,j>i}w_{ij}(\|\mathbf{x}_{t,i}\|^2+\|\mathbf{x}_{t,j}\|^2)\bigg] +\alpha\sum_{t=1}^{T-1}\|\mathbf{W}_t-\mathbf{W}_{t+1}\|_1\\
    & s.t. \quad0\leq w_{ij}\leq 1,\quad\forall i,j, \quad\sum_{i,j>i}w_{ij}=K
\end{split}
\end{equation}
where $\eta$ is a penalty coefficient, and $\mathbf{x}_{t,i}$ is the $i$-th node signal in the $t$-th time interval.
 Since the weight matrix for an undirected graph is symmetric, therefore this additional part of the new optimization can be simplified as following:
\begin{equation}
\begin{split}
    {\mathcal P} & = -2\eta\sum_{i,j>i}w_{ij}(\|\mathbf{x}_{t,i}\|^2+\|\mathbf{x}_{t,j}\|^2)\\
    & =-tr(\mathbf{x}_t^T\eta\mathcal{D}(\mathbf{W}_t)\mathbf{x_t})
\end{split}
\end{equation}
where $\mathcal{D}(\mathbf{W}_t)$ is a diagonal matrix defined above from weight matrix $\mathbf{W}_t$. Combining the two $tr(\cdot)$ expressions from Eqs.~ (3) and (4) will yield the simpler form of Eq.~(6). With a liitle more attention, one could note that this procedure naturally prefers nodes of higher energy by associating a higher weight. 
 
\section{Algorithmic Solution}
\indent The conventional use of  Lagrangian duality for solving the above  optimization model is costly in time and memory, and thus begs for an alternative.
\subsection{Projection method}
\indent To address this difficulty,  we consider the constraints as a subspace, where $\mathcal{W}$ is the whole space for graphs,  with $\mathcal{W}_{ij}\geq0$, and $W\subset\mathcal{W}$, such that $0\leq W_{ij}\leq1$. Then, we introduce a projection method for projecting the updated $W_t\in \mathcal{W}$ into the subspace $W$. Considering an updated weight matrix as a point in a high dimensional space,  we minimize the distance between the point and the subspace within the whole space by enforcing
$
    \min_{\Tilde{W}_t} \frac{1}{2} \sum_{i,j>i}(\Tilde{W}_{t,ij}-W_{t,ij})^2$, $s.t. \sum_{i,j>i}\Tilde{W}_{t,ij}=K$ and $ \Tilde{W}_t\subset \mathbf{W}
$. 
Applying the Lagrangian Duality on this minimization problem yields, \\
{\bf Claim:}
\begin{equation}
    \begin{split}
        L(\Tilde{W}_t,\kappa)= & \frac{1}{2} \sum_{i,j>i}(\Tilde{W}_{t,ij}-W_{t,ij})^2+\kappa(\sum_{i,j>i}\Tilde{W}_{t,ij}-K),\\
        & s.t. \Tilde{W}_t\subset \mathbf{W}.
    \end{split}
\end{equation}

\subsection{Proximal operator}
\indent In light of the non-smoothness of  $l_1$ norm, we call on the proximal operator to solve this part \cite{parikh2014proximal}. Firstly, the $l_1$ term $\|w_t-w_{t+1}\|_1$ in optimization (3) may be affected by the order of updating $W_t$s. Therefore, to minimize the influence of the order of updating variables, we introduce new variables $Z_t$ to replace the this term, and add a new constraint that $Z_t=W_t-W_{t+1}$. Through applying these new variables, updating $W_t$ is not influenced by the others weight matrices, and $Z_t$ gives the relaxation between each two adjacent weight matrices. In the end, this would be equivalent to the previous optimization problem, with the advantage of its decreasing the impact on the optimization caused by the order of updating variables. \\
{\bf Claim:} As a result, the Lagrangian duality form of the optimization yields the following,
\begin{equation}
    \begin{split}
        & L(W_t,X_t,\gamma,\eta,\alpha,\beta,\lambda) = \sum_{t=1}^{T} \|\mathbf{y}_t -\mathbf{x}_t\|^2\\  
        & +tr(\mathbf{x}_t^T(\gamma\mathcal{L}(\mathbf{W}_t)-\eta\mathcal{D}(\mathbf{W}_t))\mathbf{x}_t)+\alpha\sum_{t=1}^{T-1}\|Z_t\|_1\\
        & +\langle\beta_t,Z_t-W_t+W_{t+1}\rangle .
    \end{split}
\end{equation}
\indent Now we have the function of $Z_t$, denoted as $f(Z_t)=\alpha\|Z_t\|+\langle\beta_t,Z_t\rangle$, which is not a smooth but convex function over $Z_t$. To avoid smoothness point, we apply proximal operator to update $Z_t$ by projecting the point into the defined convex domain and getting closer to the optimal point. The function is defined as 
$
    \mathbf{prox}_{\lambda f}(V_t)=\operatorname*{argmin}_{Z_t}f(Z_t)+\frac{1}{2\lambda}\|Z_t-V_t\|_2^2
$
where $\lambda$ is some tuning parameter. It is clear that we achieve the optimal point $Z_t^*$, if and only if we have $Z_t^*=\mathbf{prox}_{\lambda f}(Z^*_t)$, therefore for the $k$-th iteration, we update the variable $Z_t$ as $Z^{k}_t=\mathbf{prox}_{\lambda f}(Z^{(K-1)}_t)$.

\subsection{Algorithm}
\indent The function of $X_t$ can be regarded as a convex smooth function, which allows the calculation of  the derivative of the optimization formulation over $X_t$ and  setting the value to 0.
\begin{equation}
    X^{(k)}_t = \big(\mathbf{I}+\gamma\mathcal{L}(W^{(k-1)}_t)-\eta\mathcal{D}(W^{(k-1)}_t)\big)^{-1}Y_t
\end{equation}
Since the functions of $W_t$s are smooth, we use gradient descent to update each $W_t$. The whole algorithm is presented in Algorithm 1.\\
\begin{algorithm}
    \SetKwInOut{Input}{Input}
    \SetKwInOut{Output}{Output}

    \Input{$Y_t$}
    \Output{$X_t,W_t$}
    $\alpha$, $\gamma$, $\eta$, $\lambda$ and learning rate $\tau$ are pre-defined.\\
    \While{not converged}{
    Update $X_t$ by (16)\\
    $W^{(k)}_t=W^{(k-1)}_t-\tau_1\frac{\partial}{\partial W_t}L(W_t,X_t,\gamma,\eta,\alpha,\beta,\lambda)|_{W_t^{(k-1)}}$\\
    Project $W_t^{(k)}$ to the defined domain by Projection method.\\
    Update $\mathcal{L}(W_t^{(k-1)})$ and $\mathcal{D}(W_t^{(k-1)})$ by definition.\\
    Update $Z_t$ by Proximal operator.\\
    $\beta_t^{(k)}=\beta_t^{(k-1)}-\tau_2\frac{\partial}{\partial \beta_t}L(W_t,X_t,\gamma,\eta,\alpha,\beta,\lambda)|_{\beta_t^{(k-1)}}$
    }
    \caption{algorithm for dynamic graph learning}
\end{algorithm}
\begin{figure}[h]
    \centering
    \includegraphics[scale=0.4]{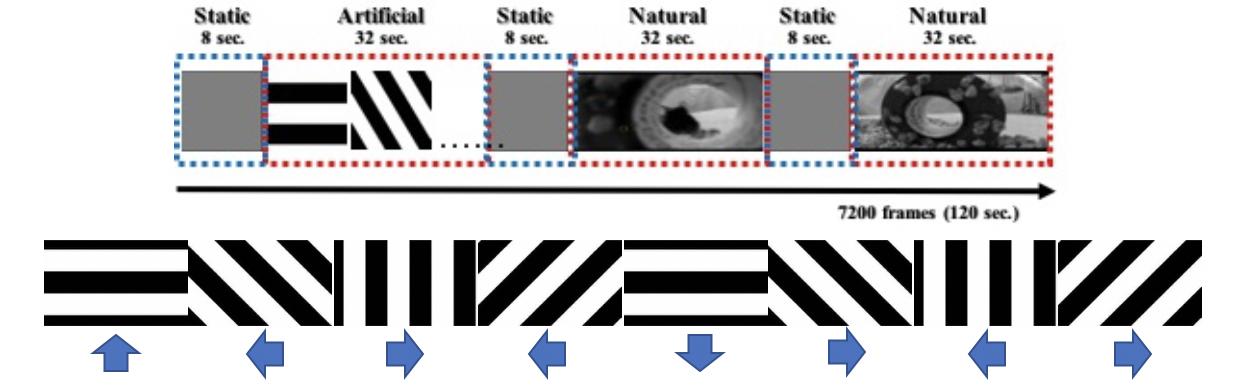}
    \caption{\textbf{Visual Stimuli:} The visual stimuli for the mouse in a single trial.}
    \label{fig:visual stimuli}
\end{figure}
\section{Experiments and results}
\indent The data in these experiments were measured in S. L. Smith's  Laboratory at  UCSB\cite{yu2018mesoscale}. They use  a two-photon electro-microscope \cite{ji2016technologies} to collect fluorescent signals. The whole experiment consists of  3 specific scenarios with a 20 trial measurement, and the stimuli in each trial are the same. The stimuli for each of the scenarios are shown in Figure 1, and consist of "gray" movie, artificial movie, natural movie 1 and natural movie 2. The dataset includes 590 neurons in V1 area and 301 neurons in AL area, and the sample rate is approximately 13.3 Hz. To select the most consistent 50  neurons in V1 area , We calculate the correlation between signals in every 2 trials for each neuron, and we choose the 50 neurons with the highest mean correlation.

\indent In addition to the similarity under similar stimuli, there is memory across the change of stimuli. The brain's reaction time for stimuli is 100ms approximately and the delay of the device is around 50 to 100ms, the time difference between 2 time points is 75ms; therefore, we choose $T=213$ in the optimization model to capture the change within 150ms, and we have 25 to 26 graphs for each stimulus. We choose $K=30$ to restrict a sparse graph, 5 percent of number of complete graph. Applying the same parameters on the signals of 20 trials, we have 8 graphs for each trial, where great variations can be observed between different trials. Therefore, we transform the weight matrix to an adjacent matrix through considering the weight matrix as the probability of connectivity between neurons, remove the edges with probability less than 0.5 and set other edges' weight to 1, then we add the adjacency matrices from different trials in the same time interval, and we choose the value of edges greater than 5.\\
\indent Through transforming the weight matrix of each graph into a vector and calculating correlation coefficient between every two vectors, e.g. the element value of $i$-th row and $j$-th column of the matrix stands for the correlation between $i$-th and $j$-th graphs' weight vectors and the matrix is symmetric obviously, the red dash lines divide the plot into small blocks, representing the exact time interval corresponding to each specific stimulus which is shown on the left of the plot, and Fig. 2 gives an intuitive view on the memories between consecutive stimuli and similarities of graphs activated by similar stimuli.\\
\begin{figure}[h]
    \centering
    \includegraphics[width=7.5cm,height=4cm]{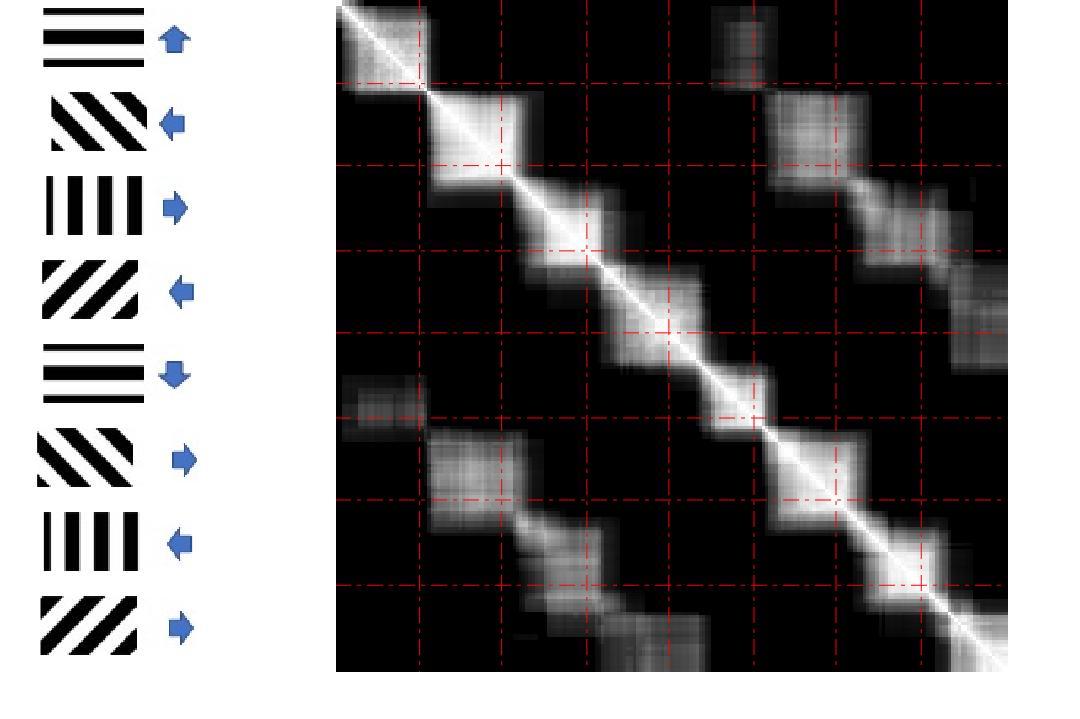}
    \caption{\textbf{Correlations between graphs}}
    \label{fig:Mrmorial graphs}
\end{figure}

\indent From this neuron signal dataset, we observe variations of neurons' connectivity over trials, but it preserves similar patterns for similar stimuli in V1 area. Through looking into different time scales, we also show the memorial patterns from one stimulus across to another. These observations can be seen as a basic step for studying brains' functional connectivity reflecting to the specific stimuli.\\

\section{Conclusion}
\indent This paper introduces an optimization model for learning dynamics of sparse graphs with sparse signals based on the graph Laplacian and its smoothness assumption without prior knowledge on the signals. Through applying three alternative solution methods, this model learns a single graph in a short time interval and a set of graphs over the whole signals, and it can capture the small change of graphs in a brief  time interval. In the experiment, we solve the difficulty of the low sample rate for detecting graphs, and discover the functional connectivity on specific stimuli instead of revealing the physical connections of neurons. More future research should focus on discovering brains with more datasets and measuring methods, which will support future discovery on understanding how neurons collaborate with each other and how brains work. A future plan is to optimize the model to  learn the transformation of graphs.



%

\nocite{*}
\bibliographystyle{IEEEtran}
\bibliography{reference.bib}

\end{document}